\documentclass[
  reprint,            % two-column, close to journal look; use 'preprint' for single-column draft
  aps,                % publisher style (aps) -- change to 'aip' if needed
  prl,                % journal (pra, prb, prl, rmp, etc.) -- choose as appropriate
  showkeys,           % show \keywords
  superscriptaddress, % author affiliations as superscripts
  longbibliography,   % include article titles in bibliography (optional)
  floatfix            % prevent stuck floats (REVTeX recommendation)
]{revtex4-2}

\usepackage[T1]{fontenc}
\usepackage[utf8]{inputenc} % Overleaf default; safe to keep
\usepackage{lmodern}        % scalable font

\DeclareUnicodeCharacter{2212}{-}

\usepackage{amsmath,amssymb,amsthm} % essential math packages
\usepackage{mathtools}              % extensions to amsmath
\usepackage{bm}                     % bold math (bm)
\usepackage{braket}                 % Dirac bra-ket notation
\usepackage{siunitx}                % units and number formatting
\usepackage{graphicx}   % includegraphics
\usepackage{array}      % better table column control
\usepackage{booktabs}   % nicer tables (\toprule etc.)
\usepackage{dcolumn}    % alignment on decimal point
\usepackage{multirow}   % multi-row table cells
\usepackage{microtype}  % better typography (pdfLaTeX)
\usepackage[normalem]{ulem}
\usepackage{xcolor}
\definecolor{linkblue}{RGB}{6,69,173}
\usepackage{hyperref}
\hypersetup{
  colorlinks=true,
  citecolor=linkblue,
  linkcolor=linkblue,
  urlcolor=linkblue,
  pdfauthor={Your Name},
  pdftitle={Paper Title},
  pdfsubject={Subject}
}

\usepackage[capitalise]{cleveref} % smart cross-references
\usepackage{comment}    % comment-out large blocks

\DeclareSIUnit{\eVperc}{eV\%}

\allowdisplaybreaks

\usepackage{color}
\newcommand{\red}[1]{{ #1 }}%{{ \color{red} #1 }}

\usepackage{comment}

\begin{document}

\title{Unmasking Hidden Wigner's Symmetry from First Principles}

\author{Phong Dang}
\email{pdang5@lsu.edu}
\affiliation{Department of Physics and Astronomy, Louisiana State University, Baton Rouge, LA 70803, USA}
%\affiliation{Quantum CodeX, Baton Rouge, LA, USA}

\author{Daniel Langr}
\affiliation{Department of Computer Systems, Faculty of Information Technology, Czech Technical University, Prague 16000, Czech Republic}
\affiliation{Quantum CodeX, Baton Rouge, LA, USA}

\author{Tom\'{a}\v{s}  Dytrych}
\affiliation{Department of Physics and Astronomy, Louisiana State University, Baton Rouge, LA 70803, USA}
\affiliation{Quantum CodeX, Baton Rouge, LA, USA}
\affiliation{Nuclear Physics Institute, Academy of Sciences of the Czech Republic, \v{R}e\v{z}, 25068, Czech Republic}

\author{Jerry P. Draayer}
\affiliation{Department of Physics and Astronomy, Louisiana State University, Baton Rouge, LA 70803, USA}
\affiliation{Quantum CodeX, Baton Rouge, LA, USA}

\author{David Kekejian}
\affiliation{Quantum CodeX, Baton Rouge, LA, USA}

%\date{\today}

\begin{abstract}
We present quantitative evidence that high-quality internucleon forces derived from $\chi$EFT exhibit a striking dominance of Wigner's %\violet{SU(4)} PD: redundant \violet{Since this is a PRL, mentioning SU(4) in the abstract or title if you prefer will expose your paper to a wider audience. More specifically, If I google scholar SU(4), your paper is more likely to pop if it has SU(4) in the title or abstract} \magenta{I don't want to add it for 4 reasons: i) search engines nowadays are powered by AI, they are not just keyword matchings but also content-driven, ii) SU(4) here is confusing since the entire paper uses U(4), iii) it's a PRL, every word has to matter by providing more meaning and SU(4) doesn't provide extra info, iv) I want to avoid bringing up group theory already in the abstract.} \violet{I believe google scholar still uses keyword matching when searching. You have the final decision though and is up to you. RESOLVED!}
supermultiplet symmetry, without invoking the large-$N_c$ limit of QCD or assumptions about specific nuclei. We trace the manifestation of this symmetry in nuclear structure using the \textit{ab initio} Symmetry Adapted Model (SAM) and identify suppressed spin-isospin polarizability. Our calculations show that a majority of $\rm ^4He$, $\rm ^6Li$, and $\rm ^6He$ wave functions is concentrated in a few $\rm U(4)$ irreducible representations, without imposing any \textit{a priori} constraints on the model space. This emergent feature points to a strategy for reducing explosive many-body bases of the NCSM while retaining physically important configurations needed to compute observables.%\footnote{Please use colors for your changes: \red{DL}, \green{TD}, \violet{DK+JD}}
\end{abstract}


\maketitle

\begin{figure}[h]
  \centering
  \includegraphics[width=0.34\textwidth]{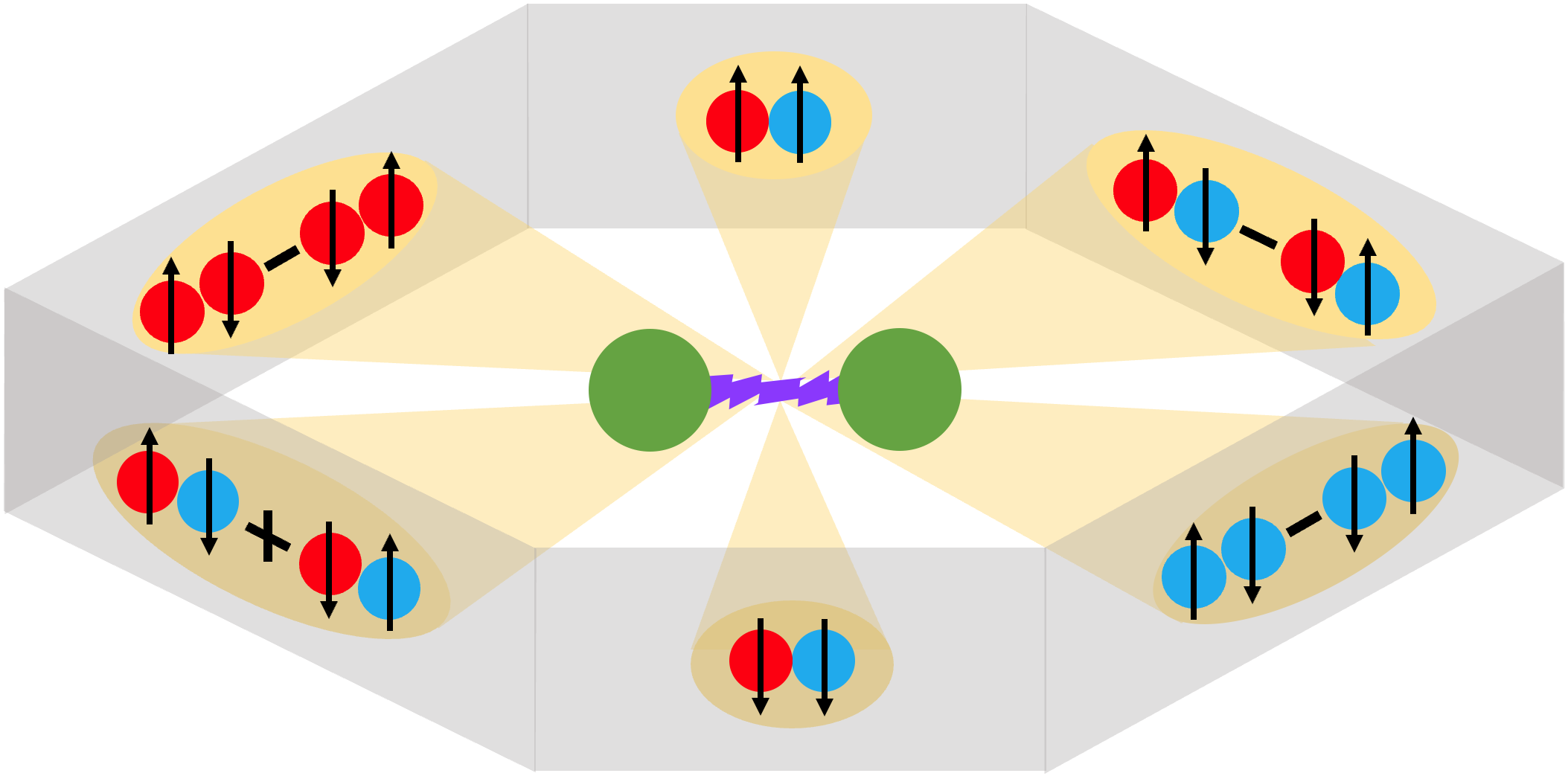}
  \caption{Illustration of Wigner's symmetry for the two-nucleon system (green balls). Neutron (red balls) and proton (blue balls) can assume spin up or down. There are two spin-singlet states for nn and pp cases. For np case, there are three spin-triplet states and one spin-singlet state. At the exact $\rm SU(4)$ limit, all six states are identical as seen by the nuclear force.}
  \label{fig:2nucleons}
\end{figure}

Symmetry is a fundamental organizing principle that appears throughout the natural and human-made world. %, from microscopic biological structures to large-scale architecture. %, symmetric patterns reflect regularities that govern complex behaviors. 
Atomic nuclei are no exception; 
%State-of-the-art \textit{ab initio} nuclear theory \cite{Hergert2020FP} has shown that, 
despite the intricacy of the nuclear force, nuclei exhibit an orderly pattern of large deformation underpinned by the recurring dominance of Elliott's $\rm SU(3)$ and symplectic $\rm Sp(3,R)$ symmetries \cite{Dytrych2007PRL,Dytrych2013PRL,Dytrych2020PRL,Launey2016PPNP,McCoy2020PRL}. In addition, nuclei display a so-called supermultiplet symmetry, an idea suggested by Wigner almost a century ago \cite{Wigner1937PR} as an \textit{umbrella} symmetry encompassing the spin and isospin degrees of freedom of nucleons via a special unitary group chain $\rm SU(4) \supset SU_S(2) \otimes SU_T(2)$. %, where the subgroups $\rm SU_S(2)$ and $\rm SU_T(2)$ manage spin and isospin degrees of freedom, respectively.
Figure \ref{fig:2nucleons} illustrates the limit of the $\rm SU(4)$ symmetry, where the six $s$-wave spin-isospin states of the two-nucleon system transform under an $\rm SU(4)$ sextet \cite{Chen2004PRL}.

%enabled accurate calculations of nuclear properties with high predictive power and quantified uncertainty, while shedding new light on fundamental understanding of the nuclear force \cite{Becker2026PRL,Sun2025PRX} and the emergence of collective phenomena. Specifically, first-principle calculations from the \textit{ab initio} Symmetry-Adapted No-Core Shell Model (SA-NCSM)  have demonstrated that the prevalence of the spatial symmetries, Elliott's $\rm SU(3)$ and symplectic $\rm Sp(3,R)$, arises naturally from realistic high-fidelity nuclear interactions without any \textit{a priory} constraints, which can be taken advantage of to control the explosive many-body basis of the NCSM via selection of configurations of high physical importance.

%SpNCCI \cite{McCoy2020PRL}

%\red{Roughly 20 years before Elliott's proposal of the $\rm SU(3)$ symmetry, Wigner suggested the existence of  ; it is known as the supermultiplet symmetry, or $\rm SU(4)$, or $\rm U(4)$ symmetry. }

Much progress has been made to establish the origin of Wigner's symmetry and its relevance to nuclear systems. Evidence for the symmetry appears in the weak binding of the deuteron \cite{Kaplan1996PLB} and nucleon-nucleon scattering phase shifts \cite{Lee2021PRL}, while its consequences extend to heavier nuclei through nuclear mass \cite{Franzini1963PL,Cauvin1981NPA}, $\beta$-decay \cite{Wigner1939PR,Fujita1965NP,Elliott1969}, $\beta\beta$-decay \cite{Vogel1986PRL,Vogel1988PLB}, electron scattering \cite{Donnelly1970AoP}, and Gamow-Teller resonances \cite{Gaponov2010PoAN}. Effective field theories (EFTs) and lattice simulations \cite{Kaplan1996PLB,Kaplan1997PRC,Cordon2008PRC,Lee2021PRL} show that the symmetry emerges in the large-$N_c$ (number of colors) limit of quantum chromodynamics (QCD) at momentum resolution around 500 MeV and in the limit of large scattering lengths \cite{Mehen1999PRL}. Wigner's symmetry also arises at the $\rm SU(3)$-flavor-symmetric point of QCD \cite{Beane2013PRC}, and is linked to entanglement suppression of the strong interaction in the infrared \cite{Beane2019PRL}. Early investigations tested the symmetry in sd-shell \cite{Hecht1974NPA,Haq1974NPA} and fp-shell nuclei \cite{Vogel1993PRC}, while algebraic solutions have been proposed to mitigate $\rm SU(4)$-symmetry breaking in heavy systems \cite{Isacker2023Symmetry,Kota2024PScr}. \red{Recently, the first \textit{ab initio} probe of the symmetry and its impact on the two-body weak current have been established via the No-Core Shell Model (NCSM) \cite{Muli2026PRL}.}

%\violet{Comment: Adding a bit more comment regarding the results of these mentioned studies will show the relevance of SU4 and will show our deep understanding of the symmetry in the broader field like you do in the next paragraph}\magenta{Please suggest your comments.} 
%\violet{I haven't read these papers. I am assuming you did or at least are familiar with their results since you included them as a reference. I am asking you to please include a sentence or two regarding them in the above paragraph} \magenta{I don't think it's necessary. It's a PRL, supposed to be short and concise, not a review paper. Also, the first sentence of the paragraph already summarizes the results.}\violet{Up to you. Worst case, if we get comments from the reviewers you can always add them later. RESOLVE as you see fit}

Nuclear physicists have strived to exploit Wigner's symmetry in first-principles modeling of atomic nuclei. Notable application is the use of $\rm SU(4)$-symmetric interactions in lattice simulations of nuclei and nuclear matter \cite{Lu2019PLB,Lu2020PRL,Shen2021EPJA,Ren2024PLB,Ren2025PRL}. The advantage of such symmetry-preserving potentials lies in their capability to mitigate the sign problem of fermionic quantum Monte Carlo \cite{Niu2025PRL} and map out nuclear structure with resolution on par with high-fidelity realistic interactions \cite{Shen2023NatComm,Shen2025PRL}. Nonetheless, understanding how the symmetry emerges in structure of nuclei from the strong nuclear force and, more importantly, leveraging the symmetry in modeling nuclear states remain elusive.

This Letter presents the first \textit{ab initio} calculation that fully employs Wigner's symmetry as a guiding principle, enabled by the \textit{ab initio} Symmetry Adapted Model (SAM), the successor to the Symmetry-Adapted No-Core Shell Model (SA-NCSM) \cite{Draayer2024PScr}. The SAM offers unequivocal evidence for the prevalence of Wigner's symmetry in realistic nuclear forces without considering the large-$N_c$ limit or any nuclei. %, and for its manifestation in nuclear structure. 
Moreover, analysis of SAM nuclear wave functions reveals a preference toward suppressed spin-isospin polarizability that complements dominant quadrupole collectivity. These wave functions are high-resolution maps of all $\rm U(4)$ irreducible representations (irreps) realized by a nucleus, made possible by the first fully generic calculator of $\rm U(4)$ coupling coefficients \cite{Dang2024EPJP2} together with its modern $\rm U(3)$ counterpart \cite{Dang2024EPJP1}. (Though $\rm SU(4)$ irreps suffice for spin and isospin decomposition, we build the SAM with $\rm U(4)$ irreps to implicitly include the particle number.) We further show that this pattern can be leveraged to compress NCSM bases to sets of physically relevant configurations that are adequate to compute nuclear observables, including binding energies, radii, electromagnetic moments and weak decay. Our long-term goal is to design a selection strategy optimized for $\rm SU(4)$-driven electroweak processes, especially $\beta$-decay for tests of the Standard Model \cite{Sargsyan2022PRL,Burkey2022PRL,King2025}. %, which are strongly driven by $\rm SU(4)$ generators.

%\red{Thanks to the advanced algebraic technologies, the SAM encodes Wigner's symmetry directly in many-body basis to represent nuclear configurations. Therefore, as it is shown in this letter, it can map out the finest details of how the symmetry manifests itself in the structure of atomic nuclei. Results show a striking evidence for the dominance of the symmetry and pave the way for a new strategy to control the model space explosion of the NCSM, while keeping track of the most relevant physics.}

%\red{Need a strong statement on why SAM? How is it more advantagous than SA-NCSM? Focus on its capability to retain configurations important for observables that mix spin and isospin simultaniously, e.g., $\beta$ decay.}

%\section{Interaction}

\begin{figure}[b!]
    \centering
    \includegraphics[width=\linewidth]{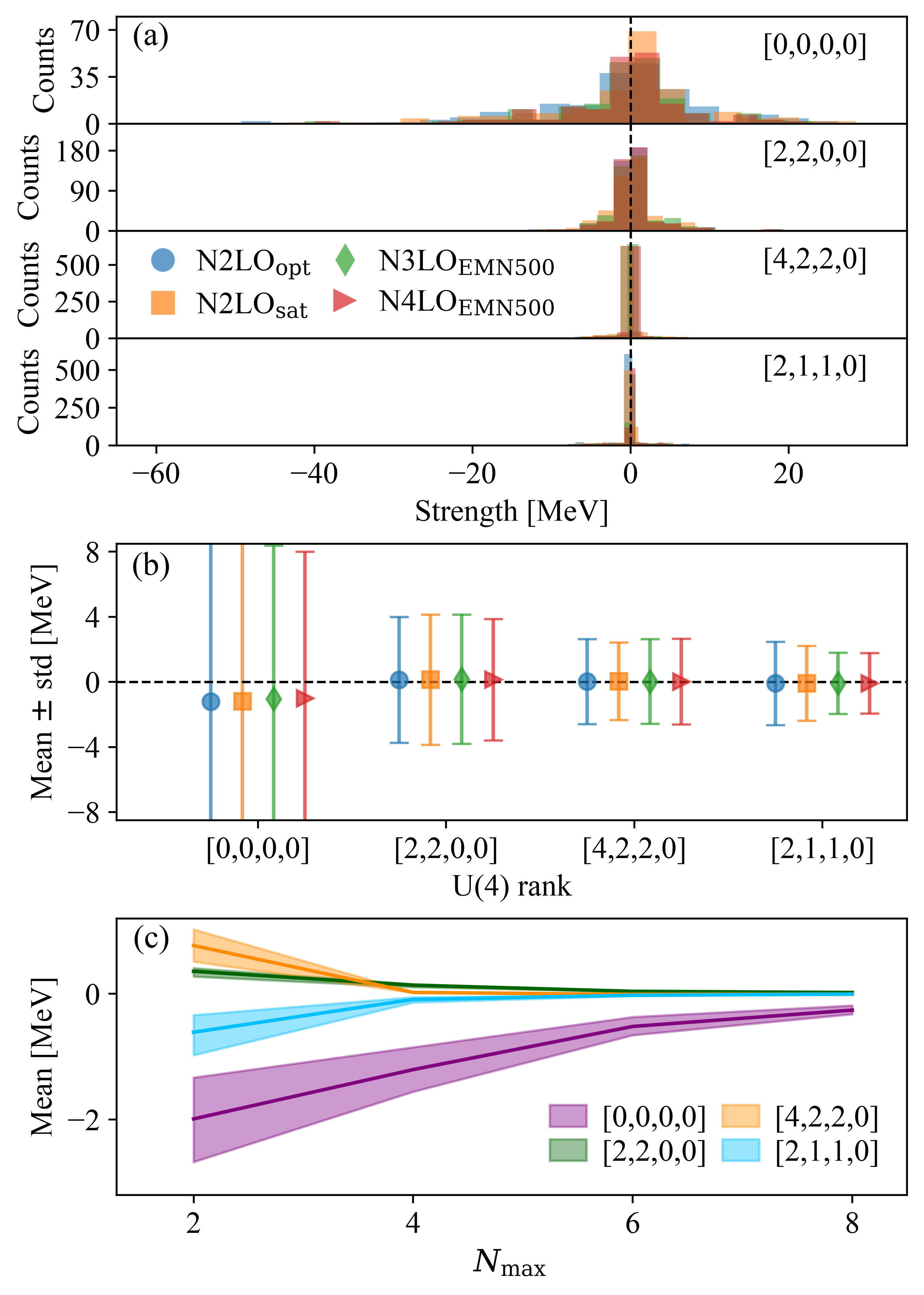}
    \caption{Expansion of four interactions, namely $\rm N2LO_{opt}$ (blue circle), $\rm N2LO_{sat}$ (gold square), $\rm N3LO_{EMN500}$ (green diamond) and $\rm N4LO_{EMN500}$ (red right triangle) %\violet{Comment: Naming the shapes are unnecessary since they are color coded and the shapes are only used in 1 of 3 pics}, PD: most PRLs do explain \violet{OK. RESOLVED!}
    in terms of $\rm U(3) \otimes U(4)$ tensors. Plot (a) shows the distributions of strengths for four $\rm U(4)$ ranks, $[0,0,0,0]$, $[2,2,0,0]$, $[4,2,2,0]$ and $[2,1,1,0]$, at $\hbar\Omega=20$ MeV and $N_{\max}=4$. Plot (b) juxtaposes %\violet{juxtaposes?} %PD: yes, please check dictionary \violet{My concern is that the word is uncommon. Maybe use the word Merge instead? Up to you though}}
    their statistical means and standard deviations (std). Plot (c) demonstrates the $N_{\max}$ dependence of the mean strengths for $\rm N2LO_{opt}$; the middle line represents $\hbar\Omega=20$ MeV, whereas the colored bands are bounded by $\hbar\Omega=15$ MeV and $\hbar\Omega=25$ MeV.}
    \label{fig:tensors}
\end{figure}

\textit{Wigner's Symmetry of the Nuclear Force}---In the SAM, %, in principle, can admit a variety of nuclear interactions from phenomenological forces to realistic ones. In this work, we employ the latter 
nucleons are strongly interacting via realistic nuclear potentials derived from the state-of-the-art $\chi$EFT %\violet{s} PD: no plural for chiEFT, see a recent review at https://link.springer.com/chapter/10.1007/978-3-032-03191-4_2#Abs1$0 \violet{RESOLVED!}
and constrained by few-body data. %high-precision 
%\violet{My concern here is that If someone familiar with NCSM reads this and they are not using $\eta$ they might be confused.} \magenta{$\eta$ is correct, check chapter 2 of Tomas' PhD thesis. YOU'VE DELETED MY ORIGINAL SENTENCE!!!} 
In conventional shell models like the NCSM, one must construct these interactions with the standard quantum numbers of the three dimensional harmonic oscillator, $\ket{\eta lj m_j m_t}$. However, as the SAM builds on a symmetry-adapted basis of the harmonic oscillator, adoption of these interactions necessitates a transformation to tensors that carry the same quantum numbers as the basis. In other words, these interactions are expanded in terms of $\rm U(3) \otimes U(4)$ tensors, each equipped with their own strength \cite{Oberhuber2021DCDS} that is determined by matrix elements of the interaction together with $\rm U(4)$ \cite{Dang2024EPJP2} and $\rm U(3)$ \cite{Dang2024EPJP1} coupling coefficients, \red{see Eqs. (1)-(2) in the Supplemental Material (SM) \cite{Supplemental}}. While facilitating the admittance of high-fidelity interactions, %into a symmetry-adapted basis, this task, though very computationally expensive, 
this task can reveal special symmetries hidden in nuclear potentials, which might not be trivial to realize in other bases.% in the traditional harmonic oscillator basis.

We explore the two-body part of four $\chi$EFT interactions, namely $\rm N2LO_{opt}$ \cite{Ekstrom2013PRL}, $\rm N2LO_{sat}$ \cite{Ekstrom2015PRC}, $\rm N3LO_{EMN500}$ and $\rm N4LO_{EMN500}$ by Entem-Machleidt-Nosyk with a $500$ MeV regulator cutoff \cite{Entem2017PRC}, whose matrix elements are streamlined from \verb|NuHamil| library \cite{Miyagi2023EPJA}. %\violet{Comment: Are you trying to say we are examining these potentials without any limitations and/or assumptions regarding nuclei, Nmax, etc. If yes, your next statement does not convey that!} PD: I've edited for clarification. \violet{Can we please say?} \magenta{OK, slightly edited.}\violet{RESOLVED!} 
Without limiting to any specific nuclei or renormalization, we study these potentials across various shell separation energies, $\hbar\Omega$, and $N_{\max}$ model spaces in which the tensor expansion is capped (definition of $N_{\max}$ herein is analogous to that of the NCSM). There are four $\rm U(4)$ ranks in the expansion, $[0,0,0,0]$, $[2,2,0,0]$, $[4,2,2,0]$ and $[2,1,1,0]$, as shown in Fig. \ref{fig:tensors}. The first is %\violet{a} PD: I use it as an adjective, not a noun \violet{Right! RESOLVED!}
$\rm U(4)$-scalar and representative of the central nuclear force, whereas the rest appear due to the symmetry-breaking interactions (e.g., spin-orbit, tensor, Coulomb). To gain quantitative insights into the signature of Wigner's symmetry, we perform statistical analysis of the strength distributions in the tensor expansion.% of these four nuclear potentials.

%\magenta{The next 2 paragraphs are a bit broken because of the figure, once I clean up edits, it will be okay.}\violet{RESOLVED!}

Our probe reveals that the largest strengths %responsible for nuclear binding 
are concentrated in $\rm U(4)$-scalar tensors, Fig.\hyperref[fig:tensors]{~\ref*{fig:tensors}(a)}, which in turn 
%\violet{Comment: I believe you mean to say preserve and not protect;} PD: I do mean protect, check Refs. [21] and [27]\violet{I did and I beleive it should be preserve. Protect implies the U(4) scalar tensors protect spin-isospin content against other types of tensors whereas preserve means they keep the spin-isospin content unchanged. Isn't it the latter?}\magenta{If you think of only tensor characters, then ``preserve'' suffices. However, ``protect'' is stronger word here, since the large strengths of the scalar tensors in fact diminish the mixing generated by non-scalar tensors. If the large strengths were concentrated in the other tensors, spin-isospin content would be strongly mixed.}\violet{Good justification. RESOLVED. I am good with 'protect'}
protect the spin-isospin content of wave functions. This is further supported by evidence that the standard deviations of the strength distributions of the scalar tensors are roughly 2-5 times larger than those of the other $\rm U(4)$ tensors across the four interactions,
%; and despite larger counts of non-scalar tensors, ...\% of them are resided within $2\sigma$
Fig.\hyperref[fig:tensors]{~\ref*{fig:tensors}(b)}. Moreover, the mean strengths of the scalar tensors dominate by a significant departure from zero; indeed, at $N_{\max}=4$, they are about 6-9, 12-14, 28-53 times larger than those of $[2,2,0,0]$, $[2,1,1,0]$ and $[4,2,2,0]$ tensors, respectively. 
% \violet{Comment: This should come after *** but breaks the page for some reason. Please fix!} PD: don't worry about it, I will fix it in the end\violet{RESOLVED!}
Such a dominance continues to be valid for different $\hbar\Omega$ energies and even gets enhanced in larger $N_{\max}$ spaces, Fig.\hyperref[fig:tensors]{~\ref*{fig:tensors}(c)}, notwithstanding the convergence of all mean strengths toward zero, see Fig. 1 in SM \cite{Supplemental}. These results confirm that the primary drivers of nuclear binding possess dominant $\rm U(4)$ symmetry \cite{Lu2019PLB}.

%\violet{Comment: This paragraph is a one very long sentence and therefore gets confusing towards the end. Please edit!} PD: Done.
Note that comparing the means and standard deviations of the strength distributions between these potentials reveals %\violet{reveal} PD: 'reveals' because its subject is comparison \violet{RESOLVED!}
that the most $\rm U(4)$-symmetric one is $\rm N2LO_{opt}$, which points to the softness of this interaction and suggests emergent symmetries and other large-$N_c$ constraints \cite{Lee2021PRL} be considered in optimizing realistic interactions.

\newpage
\onecolumngrid
\begin{figure}
\centering
\begin{minipage}{\textwidth}
\includegraphics[width=0.9\textwidth]{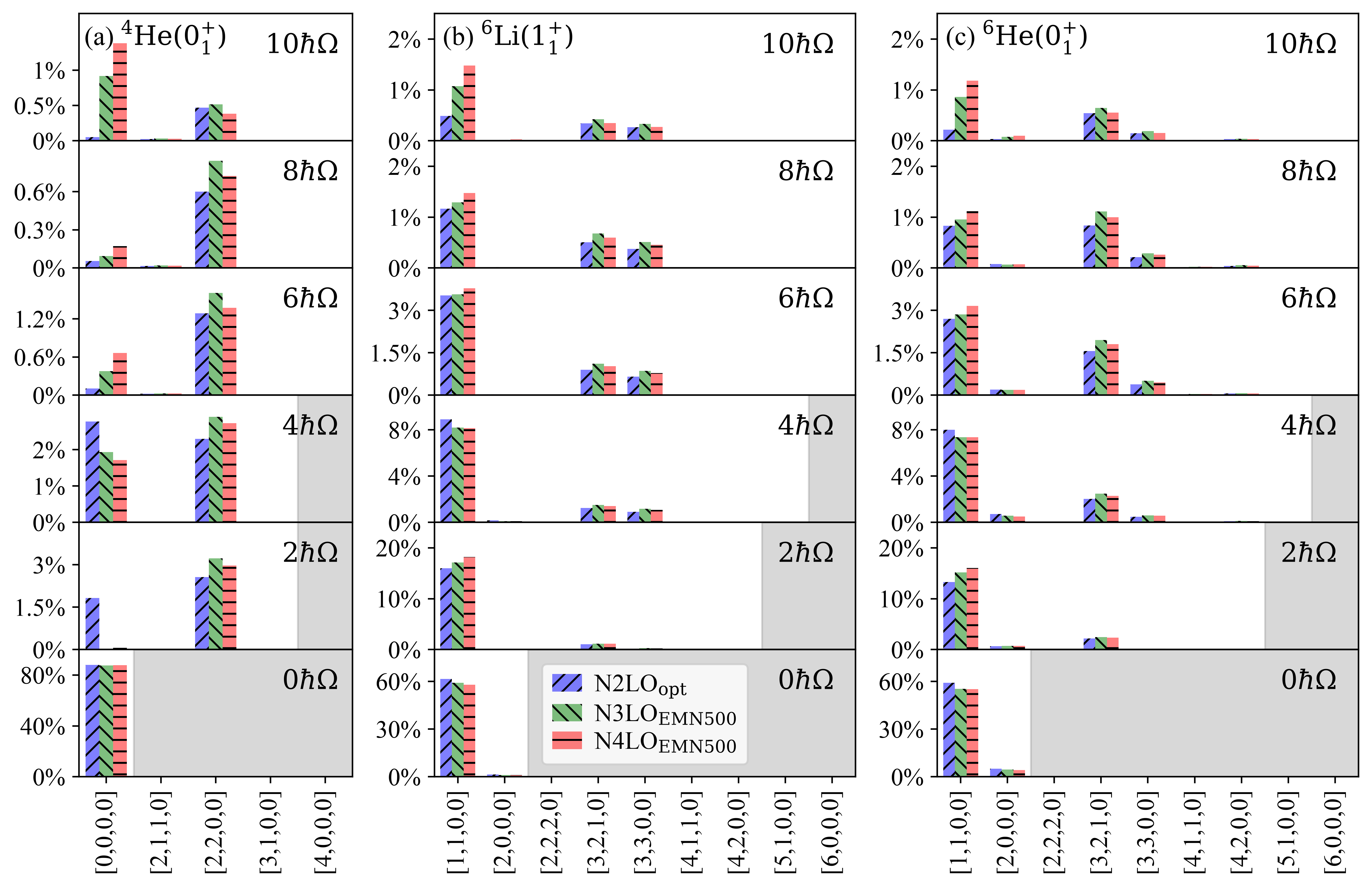}
\caption{Probability distributions of all $\rm U(4)$ irreps in the ground states of $\rm ^4He$, $\rm ^6Li$ and $\rm ^6He$ computed in \red{$N_{\max}=10$} model spaces with three bare interactions, namely $\rm N2LO_{opt}$, $\rm N3LO_{EMN500}$ and $\rm N4LO_{EMN500}$ at $\hbar\Omega=20$ MeV. The gray-shaded areas indicate that the irreps are not allowed within a given subspace. The irreps are organized in the increasing order of the $\rm SU(4)$ quadratic Casimir invariant from left to right on the horizontal axes. Here the irreps are \textit{normalized} according to \cite{Dang2024EPJP2}.}
\label{fig:wavefunctions}
\end{minipage}
\end{figure}
\twocolumngrid

%It is worth noting that $N_{\max}=0$ space represents strong repulsion of nuclear force at very short distances on the grounds that the mean strength of scalar tensors is $11.995$ MeV for $\rm N2LO_{opt}$ at $\hbar\Omega=20$ MeV, whereas those of $[2,2,0,0]$ tensors and $[2,1,1,0]$ tensors are about $-0.3$ MeV only. (There is no tensor of rank $[4,2,2,0]$ in this space.) In addition, one can observe that the mean strengths of all tensor ranks approach zero with increasing $N_{\max}$, Fig.\hyperref[fig:tensors]{~\ref*{fig:tensors}(c)}, hinting at the short-range nature of the nuclear force. Combining these two features, we identify that the $N_{\max}$ dependence of the mean strength of scalar tensors follows the behavior of the central part of the nuclear potential that is contingent upon only spatial degrees of freedom and neglects spin-isospin ones.

%\section{Structure}

\textit{Wigner's Symmetry in Nuclear Structure}---Having seen the approximate $\rm U(4)$ symmetry of nuclear interactions, one wonders how such a feature is manifested in nuclear structure. First-principles SAM calculations \red{confirm results in \cite{Muli2026PRL}} that a large fraction of nuclear wave functions is concentrated within a few irreps regardless of the interaction \red{and uncover that their dominance is persistent across each $\hbar\Omega$ subspace. Distinct from \cite{Muli2026PRL}, which extracts irrep probabilities by diagonalizing the $\rm SU(4)$ quadratic Casimir operator, $\mathcal{C}_2$, our basis is constructed explicitly on $\rm U(4)$ irreps, enabling direct reading of the irrep contribution.} We compute the ground state ($0^+_1$) of $\rm ^4He$ (even-even), the ground state ($1^+_1$) and three low-lying excited states ($3^+_1$, $2^+_1$, $1^+_2$) of $\rm ^6Li$ (odd-odd), and the ground-state rotational band ($0^+_1$, $2^+_1$) of $\rm ^6He$ (halo), followed by an analysis of all $\rm U(4)$ irrep probabilitites in their \red{$N_{\max}=10$} model spaces. The two-body Hamiltonian consists of three terms, the intrinsic kinetic energy, a realistic potential ($\rm N2LO_{opt}$, $\rm N3LO_{EMN500}$ or $\rm N4LO_{EMN500}$) and the Gloeckner-Lawson term for removal of the center-of-mass excitations \cite{Gloeckner1974PLB}. \red{Inclusion of three-body forces increases the symmetry breaking slightly but leaves the dominance pattern intact \cite{Muli2026PRL}.}

First, we observe that for all interactions a major probability of the ground-state wave functions is occupied by a single irrep residing in the $0\hbar\Omega$ subspace, Fig. \ref{fig:wavefunctions}, which has the lowest value of $\mathcal{C}_2$. %(Probablity distribution of $\rm SU(3)$ irreps in nuclear wave functions exhibit a similar feature \cite{Dytrych2013PRL}.) 
Since $\mathcal{C}_2 = S^2 + T^2 + GT^2$ \cite{Pan2023EPJP}, where the $\rm SU(4)$ generators $S$, $T$ and $GT$ are the spin, isospin and Gamow-Teller operators, respectively, such a strong preference toward low values of $\mathcal{C}_2$ may be understood as a suppression of spin-isospin polarizability of the nuclear force, which is also found to exist in infinite nuclear matter at low densities \cite{Isayev2006PRC}.

Second, as more quanta are added to excite nucleons to higher shells, more arrangements of particles are allowed, thus more irreps can enter to the model space. However, since the total number of nucleons, $A$, is constrained, at sufficiently high $N_{\max}$ where the most symmetric irrep $[A,0,0,0]$ begins to appear, no new irreps can show up. We find that in excited subspaces, mixing between different irreps takes place, Fig. \ref{fig:wavefunctions}; nevertheless, the admixture is distributed among the same set of irreps across these $\hbar\Omega$ spaces. Specifically, we witness that the ground state of the $\alpha$-particle is predominantly $\rm U(4)$-scalar ($[0,0,0,0]$) \cite{Launey2016PPNP} slightly mixed with $[2,2,0,0]$ irrep, whereas the other irreps contribute no more than 0.05\%. 
%\red{Need to link to cluster models, see \cite{Launey2016PPNP}, since we show here that the $\alpha$ particle is predominantly a U(4) scalar.} 
Similarly, the two $A=6$ isobars %, $\rm ^6Li$ and $\rm ^6He$, 
assume virtually the same set of three dominant irreps ($[1,1,0,0]$, $[3,2,1,0]$ and $[3,3,0,0]$) in each excited $\hbar\Omega$ subspace, whereas the others represent less than 0.8\%. Such akin $\rm U(4)$ structure paves the way for $\beta$-decay to occur between them \cite{Kaplan1996PLB,Muli2026PRL}. These results suggest that i) solving nuclei with small $N_{\max}$ may be sufficient to predict the most significant irreps in larger model spaces, and ii) isobaric species share analogous $\rm U(4)$ patterns allowing for electroweak processes \cite{Muli2026PRL}. 
%\red{It is worth mentioning that $\rm ^6He$ exhibits a slightly stronger mixing toward a high-$\mathcal{C}_2$ irrep, $[3,2,1,0]$, which may hint at its $\beta$-decay toward the stable nucleus $\rm ^6Li$.} %This sentence may invite questions

Third, %\violet{Comment:Earlier you didn't use the in front of SAM. We need to be consistent everywhere. Either say the SAM or just SAM} PD: I tried to use the SAM, might have missed some places.\violet{Cool! I will edit any SAM to the SAM whenever I come across. RESOLVED!}
the considered excited states demonstrate similar patterns of $\rm U(4)$ irrep probability distribution compared to the ground state (Fig. 2 in SM \cite{Supplemental}). The irrep of smallest $\mathcal{C}_2$ in the $0\hbar\Omega$ subspace remains the largest contributor; and despite a slight enhancement of admixture in excited subspaces, it is astonishingly distributed to the same set of irreps that dominate the ground state. %\magenta{\sout{Such a feature may be advantageous since one may need to solve only for the ground state to predict which irreps will play a non-negligible role in excited states.}}
%\green{I have to disagree with this statement. This is true only for the states of the ground-state band. You analyzing states dominated by the same Sp(3,R) irrep and hence they carry the same spin-isospin character as the ground state. If you look at other excited states, which are structurally not related to the ground state their U(4) content will generally be different.}\magenta{Different spins and isospins don't necessarily mean different U(4) irreps because a U(4) irrep have multiple spins and isospins; e.g., I check $0^+_1$ of 6Li, which has isospin one and it still has similar U(4) pattern. Anyway, this is definitely an open question to consider for future work. I suggest some edits above.}

%\section{Truncation}
\textit{Pattern-Recognition Nuclear Modeling}---The efficacy of the SA-NCSM stems from its exceptional capability to expose recurring orderly patterns of physically relevant degrees of freedom, %in the intricate structure of atomic nuclei, 
thus empowering a strategic selection of configurations that suffice to reproduce and predict nuclear observables accurately. Following the footstep of its predecessor, the SAM with many-body basis built explicitly on $\rm U(3)$ and $\rm U(4)$ irreps also allows for a selection scheme to refine the NCSM space based upon the dominance of these two symmetries. Specifically, SAM model spaces are defined by a pair of numbers $\langle N_{\max}^{\bot} \rangle N_{\max}^{\top}$ \cite{Dytrych2013PRL}, which implies the inclusion of all configurations up through $N_{\max}^{\bot}$ and only a subset of $\rm U(3)$ and $\rm U(4)$ irreps between $N_{\max}^{\bot}$ and $N_{\max}^{\top}$. Such a truncation still guarantees the spurious center-of-mass separation \cite{Verhaar1960NP} since the Gloeckner-Lawson operator is a $\rm U(3)$- and $\rm U(4)$-scalar and cannot connect different $\hbar\Omega$ subspaces.

\begin{figure}[t!]
  \centering
  \includegraphics[width=\linewidth]{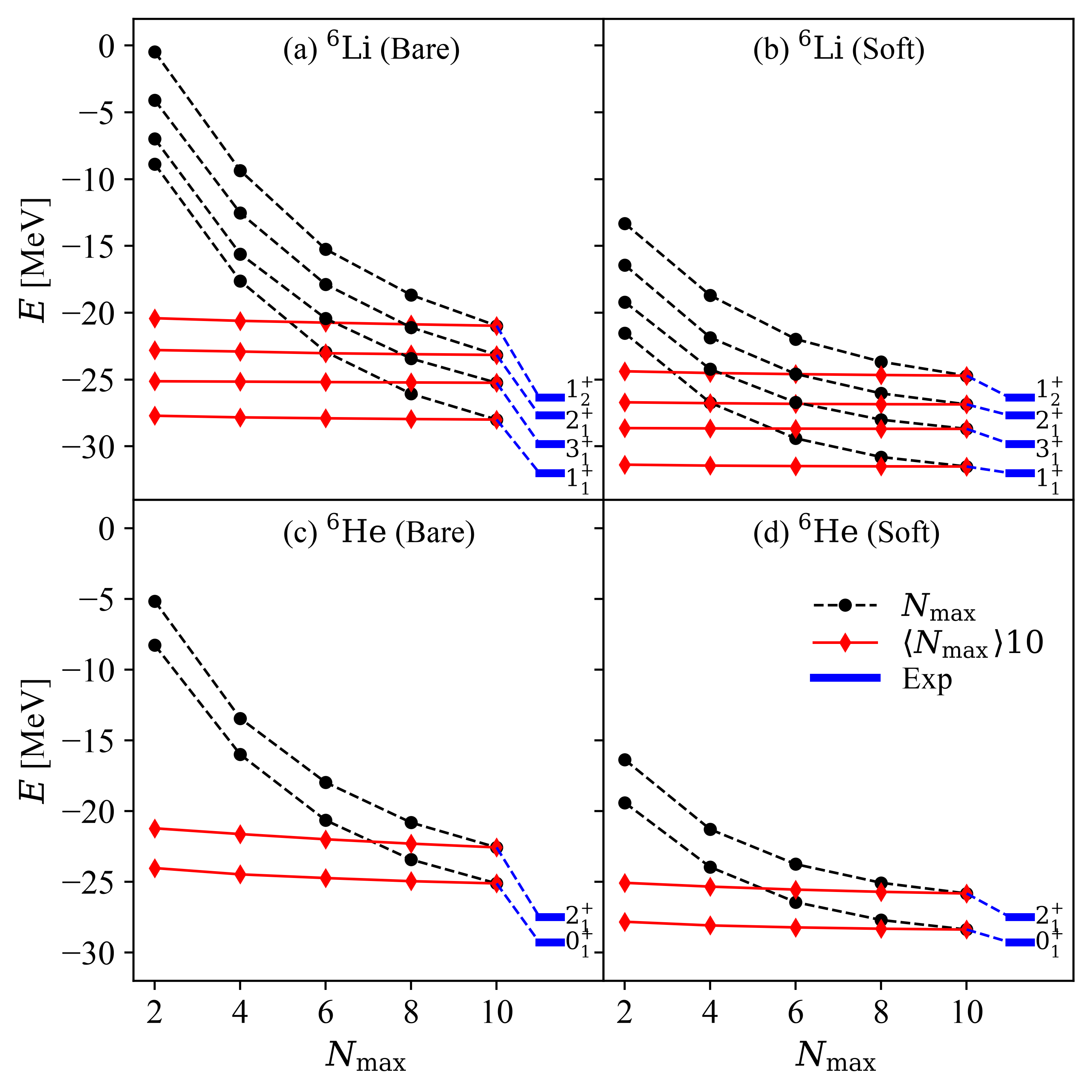}
  \caption{Binding energies for ground and excited states of $\rm ^6Li$ and $\rm ^6 He$ at $\hbar\Omega=20$ MeV. We employ $\rm N2LO_{opt}$ [(a) and (c)] and the two-body part of $\rm N3LO_{EM500}$ softened using the resolution scale $\lambda=2.0$ $\rm fm^{-1}$ [(b) and (d)]. Herein, black circles represent full model space calculations, whereas red diamonds correspond to restricted spaces with \red{$N_{\max}^{\top}=10$}. Blue horizontal bars demonstrate experimental values \cite{Tilley2002NPA}.
  }
  \label{fig:BEs}
\end{figure}

\begin{figure}[h!]
  \centering
  \includegraphics[width=\linewidth]{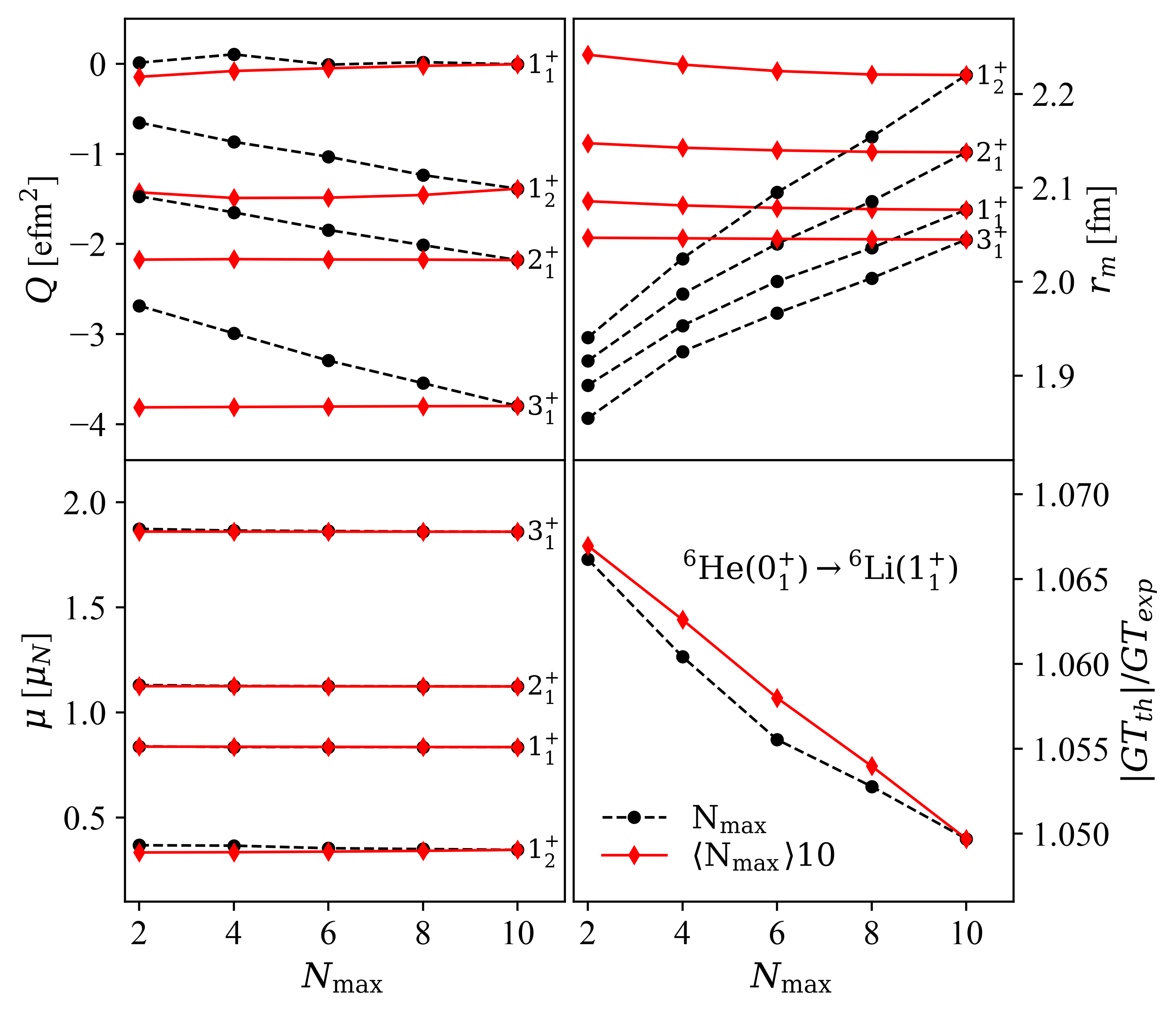}
  \caption{Selected observables of $\rm ^6Li$ using $\rm N2LO_{opt}$ at $\hbar\Omega=20$ MeV, including electric quadrupole moments $Q$, magnetic dipole moments $\mu$, point-nucleon root-mean-squared matter radii $r_m$, and reduced Gamow-Teller matrix elements $GT_{th}$ relative to the experimental value $GT_{exp}$ from \cite{Knecht2012PRC}. Herein, the markers have similar meaning to those in Fig. \ref{fig:BEs}.%; however, in the bottom right plot, the solid red line corresponds to selection of $[1,1,0,0]$, $[3,2,1,0]$ and $[3,3,0,0]$ irreps, whereas the dashed red line represents selection of $[1,1,0,0]$, $[2,0,0,0]$ and $[3,2,1,0]$ irreps.
  }
  \label{fig:obs}
\end{figure}

This Letter focuses on how $\rm U(4)$-guided selection of basis states affects \textit{ab initio} calculations of nuclear observables without any $\rm U(3)$ irrep selection as it has been investigated extensively via the SA-NCSM. %Nonetheless, we plan to employ an all-inclusive strategy in future works that fully take advantage of both the symmetries in constructing model spaces. 
%We probe the convergence of observables in restricted spaces with values calculated in complete spaces. 
Specifically, we construct model spaces for $\rm ^6Li$ and $\rm ^6He$ with \red{$N_{\max}^{\top}=10$ and vary $N_{\max}^{\bot}$ from 2 to 8}. In each $\hbar\Omega$ subspace where selection of $\rm U(4)$ irreps is performed, we keep the most dominant irreps $[1,1,0,0]$, $[3,2,1,0]$ and $[3,3,0,0]$ for both nuclei. The resulting dimensions of \red{$\langle 2 \rangle 10$, $\langle 4 \rangle 10$, $\langle 6 \rangle 10$ and $\langle 8 \rangle 10$ spaces are roughly 67-74\% for $\rm ^6Li$ and 47-57\% for $\rm ^6He$ compared with their complete $N_{\max}=10$ spaces.}
Our calculations include binding energies $E$, electric quadrupole moments $Q$, magnetic dipole moments $\mu$, point-nucleon root-mean-squared matter radii $r_m$, and Gamow-Teller $\beta$-decay reduced matrix elements $GT$, using the $\rm N2LO_{opt}$ interaction with $\hbar\Omega=20$ MeV. Among these observables, the magnetic dipole moments prove to be the least dependent on $N_{\max}$ in full and truncated spaces (deviations $\sim 10^{-3}$-$10^{-2}$ for $\rm ^6Li$, Fig. \ref{fig:obs}).

Figure \ref{fig:BEs} %\violet{Figure 4 is in the previous page. Can you please move it here?}\magenta{These organization details are messed up by the comments. I will fix once the edits are cleansed.}\violet{RESOLVED!} 
demonstrates the efficacy of the selection for computing binding energies. For the stable nucleus $\rm ^6Li$, the \red{$\langle 2 \rangle 10$} space already delivers 99\%, 99\%, \red{98\%} and \red{97\%} of binding energies for the states $1^+_1$, $3^+_1$, $2^+_1$ and $1^+_2$, respectively, relative to the full-space calculations, Fig.\hyperref[fig:BEs]{~\ref*{fig:BEs}(a)}. Meanwhile, the same level of predictions for the ground state and first excited state of $\rm ^6He$ is achieved in the \red{$\langle 6 \rangle 10$} space, Fig.\hyperref[fig:BEs]{~\ref*{fig:BEs}(c)}. Moreover, varying $N_{\max}^{\bot}$ produces almost identical results with differences in the order of $10^{-1}$ MeV. %\red{Fitting \cite{Maris2009PRC,Sargsyan2023PRC}.}

We also explore combining the same selection with a low-momentum interaction obtained from the similarity renormalization group (SRG) \cite{Bogner2010PPNP}. We use the two-body part of $\rm N3LO_{EM500}$ potential by Entem-Machleidt with 500 MeV regulator cutoff \cite{Entem2003PRC}, which is softened %\violet{Do you mean is softer or has been softened? is soften not grammatically correct} \magenta{Done.} 
using the momentum scale $\lambda=2.0$ $\rm fm^{-1}$ \cite{Jurgenson2009PRL}, and compute binding energies for the same states for $\rm ^6Li$ and $\rm ^6He$. Incorporating both methodologies is very effective: this soft interaction produces faster convergence to experiment than the $\rm N2LO_{opt}$, and the selection preserves that trend. \red{Specifically, the binding and excitation energies for both nuclei converge toward experimental values within 0.3-1.1 MeV, and the $\langle 2 \rangle 10$ space reproduces binding of $\rm ^6Li$'s ground and excited states within 330 keV of the full-space results, Fig.\hyperref[fig:BEs]{~\ref*{fig:BEs}(b)}, while 98\% and 97\% of $0^+_1$ and $2^+_1$ bindings are attained in the same space of $\rm ^6He$, Fig.\hyperref[fig:BEs]{~\ref*{fig:BEs}(d)}.}

To investigate how spin-isospin-guided selection affects long-range correlations in the wave functions, we examine the electric quadrupole moments and point-nucleon root-mean-squared matter radii of $\rm ^6Li$, Fig. \ref{fig:obs}. First, all restricted spaces reproduce the correct sign of the challenging $1^+_1$ quadrupole moment and yield nearly identical values for $3^+_1$ and $2^+_1$ relative to full-space results, while differences for $1^+_2$ remain at only a few percent. Second, the selected spaces produce nearly indistinguishable radii to those of the complete spaces, with deviations appearing only at the second decimal place for all states.

Of special interest in the SAM development is the weak process, $\beta$-decay, because the transition is dominated by the Gamow-Teller (\textit{GT}) operator \cite{Gysbers2019NatP}, an $\rm SU(4)$ generator \cite{Muli2026PRL}, implying that a basis encoding Wigner's symmetry can guide selection of configurations that capture vital spin-isospin correlations for this process. Figure \ref{fig:obs} displays the results obtained from the full and restricted model spaces. First, our complete \red{$N_{\max}=10$} reduced matrix element agrees with NCSM calculation using the same interaction \cite{Shin2017JPG} and differs from experiment within \red{5\%.} Moreover, we find that $[1,1,0,0]$ strongly dominates the process, contributing roughly 90\% of the total \textit{GT}. Second, we witness that values from the selected spaces slightly exceed complete spaces of lower $N_{\max}$, which we attribute to an enhanced $[1,1,0,0]$ contribution and will investigate other truncation choices in future work. Nevertheless, that the restricted spaces yield deviation within 2\% of the complete space demonstrates that essential configurations are retained.

%Second, we observe an intriguing phenomenon that GT values from the selected spaces exhibit a linear dependence on $N_{\max}^{\bot}$ and are higher than those from complete spaces of lower $N_{\max}$. To understand this behavior, we perform a separate calculation that keeps three irreps $[1,1,0,0]$, $[2,0,0,0]$ and $[3,2,1,0]$, which produces values below the full $N_{\max}=2,4,6$ results and closer to the complete $N_{\max}=8$ value. Thus, we believe that this phenomenon occurs due to the fact that $[3,3,0,0]$ irrep has higher $\mathcal{C}_2$ than $[2,0,0,0]$ irrep, insinuating that calculation of $\beta$-decay is sensitive to selected irreps and may prefer low-$\mathcal{C}_2$ ones.

%\section{Conclusions}
In summary, we report an exceptional predominance of Wigner's symmetry in low-energy realistic nuclear forces without examining the large-$N_c$ limit or %\violet{making any assumptions or limitations regarding any nuclei}\magenta{I think it's important not to lengthen sentences without adding any substanstial meaning. Every word counts!}\violet{if we say "making assumptions about any nucler" instead then I am good with that.}\magenta{OK} 
making assumptions about any nuclei. That we find the $\rm N2LO_{opt}$ interaction to be slightly more symmetric than the others suggests emergent symmetries and other large-$N_c$ properties %of the strong interaction %\violet{must be OR will be appreciated} \magenta{Gramma: suggest + verb-ing or suggest + that ... should + verb and in high level, should can be omitted}\violet{I prefer should to be included but it is up to you. RESOLVE as you see fit} 
be appreciated in future structure calculations. %\green{What? The end of this sentence sounds a bit cumbersome. I don't really understand what you are trying to say.}\magenta{U(4) symmetry is known to emerge in the large-Nc limit of QCD. The fact that we find N2LOopt to be more U(4)-symmetric than the others suggests that the symmetry itself and other emergent phenomena at large-Nc should be considered in structure calculations, e.g., optimizing realistic interaction. A similar statement was raised in \cite{Lee2021PRL}, where the authors pinned down an optimal momentum scale for chiEFT at 500MeV.} 
We trace the presence of the symmetry in wave functions of some light nuclei and identify a suppression of spin-isospin polarizability of the nuclear force. Recognizing the recurring pattern of the $\rm U(4)$ symmetry, we test the potency of a strategic selection that filters out less important configurations in NCSM model spaces. Calculations of nuclear observables in truncated spaces prove to retain the accuracy of our predictions. It is also seen that deployment of a low-momentum interaction in symmetry-guided bases can be powerful by preserving the fast convergence in smaller spaces. Nonetheless, much more work remains to be done, our ultimate goal is to establish a comprehensive strategy that can go upward in the nuclear chart and tackle heavy species, one that combines high-quality nuclear forces, state-of-the-art SRG methods %\violet{maybe we should say methods since there are multiple SRG approaches}\magenta{OK}\violet{RESOLVED} 
and many-body bases guided by emergent orderly patterns.

% Acknowledgement
We thank Anna McCoy, Feng Pan, Dean Lee, Bingcheng He, Thomas Papenbrock, William Shelton, Kristina D. Launey and József Cseh for useful discussions and insights. This work was supported by LSU Sponsored Research Rebate Program as well as LSU Foundation's Distinguished Research Professorship Program. PD acknowledges support from the Quad Fellowship of the Institute of International Education and from LSU Summer Funding Opportunity for Graduate Students in STEM and Business Programs. This work was performed using high performance computing resources provided by LSU (www.hpc.lsu.edu).

All data produced in this work are included in the Letter. Specific numbers of quantities that are reported figuratively can be obtained from PD upon request.

% -------------------------
% Bibliography (choose style)
% -------------------------
%\newpage
%\clearpage
% For APS journals it's common to use BibTeX:
\bibliography{References} % References.bib file in project
% If you prefer BibLaTeX or a manual .bbl, adapt accordingly.

%\begin{comment}

\newpage
\onecolumngrid
\section{Supplemental Material}

\subsection{Transformation of nuclear interactions to $\rm U(3) \otimes U(4)$ tensors}

The SA-NCSM admits realistic nuclear interactions by expanding them in terms of two-body $\rm SU(3) \otimes SU(2)$ tensors which are coupled products of creation and annihilation operators \cite{Oberhuber2021DCDS}; each tensor is equipped with a strength that is computed from normal ordered two-body $J$-coupled matrix elements \cite{Miyagi2023EPJA} along with $\rm SU(3)$ coupling and $\rm SU(2)$ recoupling coefficients. Since the SAM works in $\rm U(3) \otimes U(4)$-adapted basis, we extend this formalism to transform nuclear interactions into $\rm U(3) \otimes U(4)$ tensors, which is only possible with the first generic calculator of $\rm U(4)$ coupling coefficients \cite{Dang2024EPJP2}.

The tensor expansion of a two-body nuclear potential is computationally demanding and is obtained via the following expression
\begin{equation}
\label{eq:tensors}
  \mathcal{V} = \frac{1}{4}  
  \sum_{\tiny{\begin{array}{c}
  \eta_a \eta_b \eta_c \eta_d \\ 
  \Gamma^3_i \Gamma^4_i \Gamma^3_f \Gamma^4_f \\ 
  \rho_0 \Gamma^3_0 \kappa_0 L_0 \\
  \eta_0 \Gamma^4_0 \zeta_0 S_0 T_0
  \end{array}}} g 
  \left\{ \left[\mathbf{a}^{\dagger [\eta_a,0,0]}_{[1,0,0,0]} \times \mathbf{a}^{\dagger [\eta_b,0,0]}_{[1,0,0,0]} \right]^{\Gamma^3_f}_{\Gamma^4_f} \times \left[\tilde{\mathbf{a}}^{[\eta_d,\eta_d,0]}_{[1,1,1,0]} \times \tilde{\mathbf{a}}^{[\eta_c,\eta_c,0]}_{[1,1,1,0]} \right]^{\Gamma^3_i}_{\Gamma^4_i} \right\}^{\rho_0 \Gamma^3_0 \eta_0 \Gamma^4_0}_{\kappa_0 L_0 \zeta_0 S_0 T_0 J_0=M_{J0}=M_{T0}=0},
\end{equation}
where the shell numbers ($\eta_a$, $\eta_b$, $\eta_c$, $\eta_d$) run from 0 to the maximum shell, and $\Gamma^3$ and $\Gamma^4$ are shorthand notations for general $\rm U(3)$ and $\rm U(4)$ irreps, respectively. Note that $T_0$ is not vanishing as a generic two-body nuclear potential consists of isoscalar, isovector and isotensor parts. The strengths $g$ in the expansion are computed as follows
\begin{align}
  g = & \sum_{\kappa_i L_i \kappa_f L_f} \langle \Gamma^3_f \kappa_f L_f ; \Gamma^3_i \kappa_i L_i || \Gamma^3_0  \kappa_0 L_0 \rangle_{\rho_0} \nonumber\\
  & \sum_{l_a l_b} \langle [\eta_a,0,0] l_a ; [\eta_b,0,0] l_b || \Gamma^3_f  \kappa_f L_f \rangle \sum_{l_c l_d} \langle [\eta_d,\eta_d,0] l_d ; [\eta_c,\eta_c,0] l_c || \Gamma^3_i  \kappa_i L_i \rangle 
   \sum_{j_a j_b j_c j_d} \frac{1}{\mathcal{N}_{ab} \mathcal{N}_{cd}} \nonumber \\
   & \sum_{J T_i T_f} \langle \eta_a l_a j_a, \eta_b l_b j_b; J T_f|| \mathcal{V} || \eta_c l_c j_c, \eta_d l_d j_d; J T_i \rangle \sqrt{\frac{(2J+1)(2T_f+1)}{2T_0 + 1}} (-)^{\eta_c + \eta_d + 1 - j_c - j_d + J + T_i} \nonumber\\ 
   & \sum_{S_i S_f} U\left(\begin{array}{ccc}
  l_a & 1/2 & j_a \\ 
  l_b & 1/2 & j_b \\ 
  L_f & S_f & J 
\end{array}\right) \langle [1,0,0,0]1/2, 1/2; [1,0,0,0]1/2, 1/2 || \Gamma^4_f S_f T_f \rangle \nonumber\\ 
& U\left(\begin{array}{ccc}
  l_d & 1/2 & j_d \\ 
  l_c & 1/2 & j_c \\ 
  L_i & S_i & J 
\end{array}\right) \langle [1,1,1,0]1/2, 1/2; [1,1,1,0]1/2, 1/2 || \Gamma^4_i S_i T_i \rangle \nonumber\\
& U\left(\begin{array}{ccc}
  L_f & S_f & J \\ 
  L_i & S_i & J \\ 
  L_0 & S_0 & 0 
\end{array}\right) \langle \Gamma^4_f S_f T_f ; \Gamma^4_i S_i T_i || \Gamma^4_0 \zeta_0 S_0 T_0 \rangle_{\eta_0},
\end{align}
where
\begin{itemize}
  \item $\langle \Gamma^3_f \kappa_f L_f ; \Gamma^3_i \kappa_i L_i || \Gamma^3_0  \kappa_0 L_0 \rangle_{\rho_0}$, $\langle [\eta_a,0,0] l_a ; [\eta_b,0,0] l_b || \Gamma^3_f  \kappa_f L_f \rangle$ and \\ $\langle [\eta_d,\eta_d,0] l_d ; [\eta_c,\eta_c,0] l_c || \Gamma^3_i  \kappa_i L_i \rangle$ are $\rm U(3) \supset SO(3)$ reduced coupling coefficients \cite{Dang2024EPJP1},
  \item $\langle \Gamma^4_f S_f T_f; \Gamma^4_i S_i T_i || \Gamma^4_0 \zeta_0 S_0 T_0 \rangle_{\eta_0}$, $\langle [1,0,0,0]1/2, 1/2; [1,0,0,0]1/2, 1/2 || \Gamma^4_f S_f T_f \rangle$ and \\ $\langle [1,1,1,0]1/2, 1/2; [1,1,1,0]1/2, 1/2 || \Gamma^4_i S_i T_i \rangle$ are $\rm U(4) \supset SU_S(2) \otimes SU_T(2)$ reduced coupling coefficients \cite{Dang2024EPJP2},
  \item $U(...)$ coefficients are $\rm SU(2)$ recoupling coefficients,
  \item $\mathcal{N}_{ab} = 1 / \sqrt{1 + \delta_{\eta_a \eta_b} \delta_{l_a l_b} \delta_{j_a j_b}}$ and $\mathcal{N}_{cd} = 1/ \sqrt{1 + \delta_{\eta_c \eta_d} \delta_{l_c l_d} \delta_{j_c j_d}}$ are normalization factors,
  \item $\langle \eta_a l_a j_a, \eta_b l_b j_b; J T_f|| \mathcal{V} || \eta_c l_c j_c, \eta_d l_d j_d; J T_i \rangle$ are normalized and antisymmetrized $JT$-coupled reduced matrix elements of the potential \cite{Miyagi2023EPJA}.
\end{itemize}

\subsection{Tensor expansion statistics of $\rm N2LO_{sat}$, $\rm N3LO_{EMN500}$ and $\rm N4LO_{EMN500}$}

\begin{figure}[h!]
\centering
\begin{minipage}{\textwidth}
\includegraphics[width=0.9\textwidth]{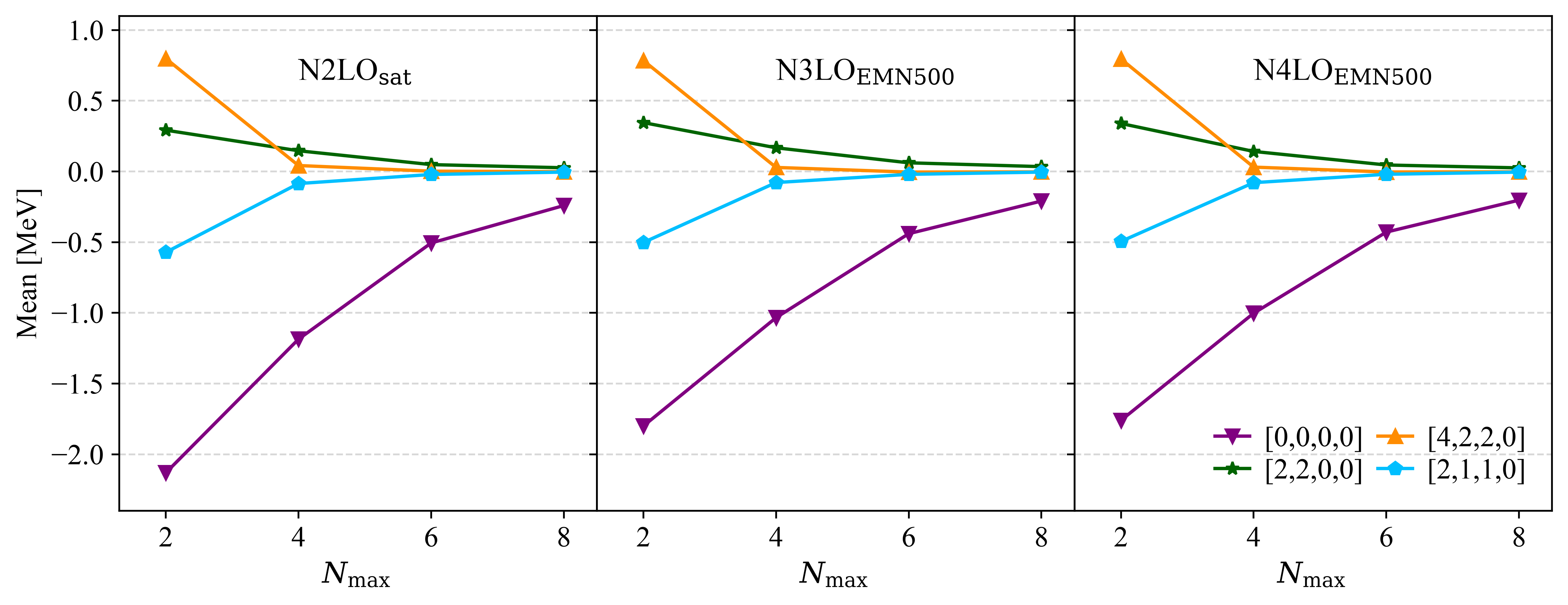}
\caption{Dependence on $N_{\max}$ of the mean strengths for four $\rm U(4)$ ranks, $[0,0,0,0]$ (purple down triangle), $[2,2,0,0]$ (green stars), $[4,2,2,0]$ (orange up triangle) and $[2,1,1,0]$ (blue pentagons) in the tensor expansion of $\rm N2LO_{sat}$, $\rm N3LO_{EMN500}$ and $\rm N4LO_{EMN500}$ at shell separation energy $\hbar\Omega=20$ MeV.}
\label{fig:U4-means}
\end{minipage}
\end{figure}

In analyzing the behavior of the mean strengths of $\rm U(4)$ tensors as one increases $N_{\max}$ for $\rm N2LO_{opt}$ (Fig. \ref{fig:tensors}) and $\rm N2LO_{sat}$, $\rm N3LO_{EMN500}$, $\rm N4LO_{EMN500}$ (Fig. \ref{fig:U4-means}), we further identify three common features across all these nuclear potentials:
\begin{enumerate}
  \item As one expands the model space of tensor expansion with higher $N_{\max}$, %\violet{Comment: Long-range tensors start to appear that connect.. }\magenta{I wrote an advanced inverse clause}\violet{I checked with chatgpt and gemini and they both said starting to appear is not the correct use. Please double check and RESOLVE} \magenta{Use Claude: ``With the full context, the inverted construction "starting to appear are..." is grammatically sound — the introductory clause anchors it properly.''} 
  starting to appear are ``long-range'' tensors that can connect configurations differing in a large number of harmonic oscillator quanta. These tensors, though large in their numbers, possess very small strengths, which in turn bring down the overall mean strengths in the tensor expansion, demonstrated by the fact that all mean strengths converge toward zero. This behavior is consistent with the short-range nature of the nuclear force.
  \item In spite of the convergence trend, at each value of $N_{\max}$, the scalar tensors always eclipse the other tensors. Specifically, at $N_{\max}=8$, the mean strengths of the $\rm U(4)$-symmetric tensors dominate those of $[2,2,0,0]$, $[2,1,1,0]$ and $[4,2,2,0]$ by a factor of 6-12, 41-59, 52-143, respectively.
  \item The mean strengths of the $[4,2,2,0]$ tensors approach zero the most quickly. %\violet{more quickly than others OR the quickest} \violet{Same here: I checked with chatgpt and gemini and seems like most quickly is formal without the and standard way is to say the quickest. Please double check and RESOLVE} \magenta{Check \href{https://forum.wordreference.com/threads/the-most-quickly.3532083/$0}{this}}. 
  It is seen that at $N_{\max}=2$, these tensors seem to be the second dominant contributor to the nuclear force, however, at $N_{\max}=4$, they begin to intrude below the $[2,2,0,0]$ tensors with the mean under 0.05, being the smallest among the four $\rm U(4)$ ranks.
\end{enumerate}

\subsection{Wigner's symmetry in excited states}

\begin{figure}[h!]
\centering
\begin{minipage}{\textwidth}
\includegraphics[width=0.9\textwidth]{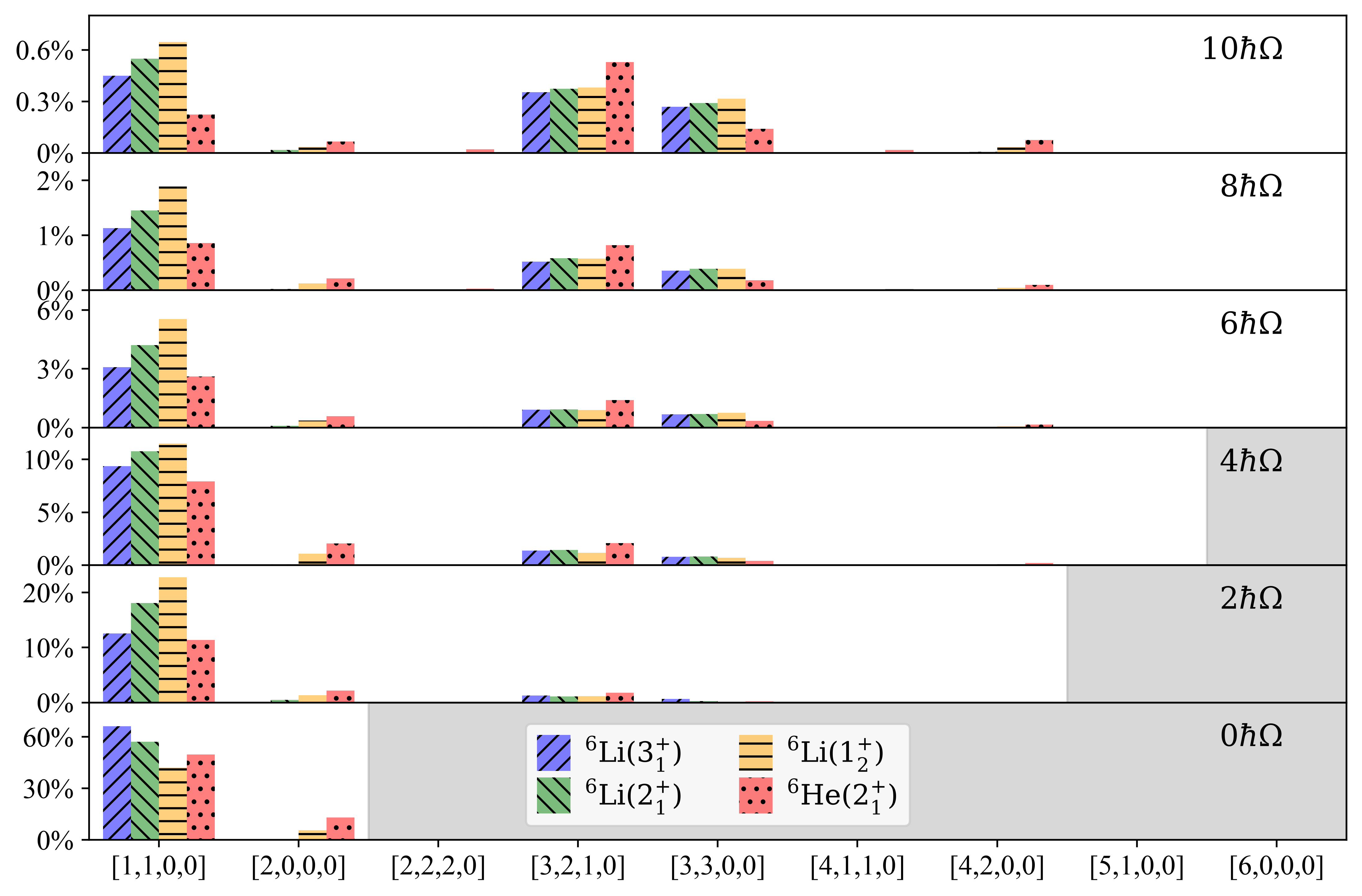}
\caption{Probability distributions of all $\rm U(4)$ irreps in low-lying excited states of $\rm ^6Li$ ($3^+_1$, $2^+_1$, $1^+_2$), and $\rm ^6He$ ($2^+_1$) computed in $N_{\max}=10$ model spaces. The gray-shaded areas indicate that the irreps are not allowed within a given subspace. The irreps are organized in the increasing order of the quadratic Casimir invariant of the $\rm SU(4)$ group from left to right on the horizontal axes. Calculations are performed with $\rm N2LO_{opt}$ at intershell distance $\hbar\Omega=20$ MeV. Here the irreps are \textit{normalized} according to \cite{Dang2024EPJP2}.}
\label{fig:excited-states}
\end{minipage}
\end{figure}

Figure \ref{fig:excited-states} maps out the probability distributions of three excited states of $\rm ^6Li$ ($3^+_1$, $2^+_1$, $1^+_2$), and the first excited state of $\rm ^6He$ ($2^+_1$) across six $\hbar\Omega$ subspaces of their $N_{\max}=10$ model spaces. It can be seen that the most prominent $\rm U(4)$ irreps in these states are $[1,1,0,0]$, $[3,2,1,0]$ and $[3,3,0,0]$, which also dominate the structure of their ground states. As pointed out in the main text, there is a slight enhancement of mixing. Specifically, the $1^+_2$ state of $\rm ^6Li$ and the $2^+_1$ state of $\rm ^6He$ see an increased presence of the $[2,0,0,0]$ irrep.

%\end{comment}

\end{document}